\documentclass[conference,a4paper]{IEEEtran}
\IEEEoverridecommandlockouts

\usepackage{amsmath,amssymb,epsfig,bbm}
\usepackage{graphicx,color}

\newtheorem{theorem}{Theorem}[section]
\newtheorem{fact}{Fact}[section]

\def\estimateG{\widehat{G}}
\def\estimateA{\widehat{A}}

\def\spacing{s_b}

\def\inas{\stackrel{a.s.}{\longrightarrow}}
\def\inltwo{\stackrel{{\cal L}^2}{\longrightarrow}}
\def\weakly{\stackrel{d}{\longrightarrow}}

\def\eE{{\mathbb E}}
\def\pP{{\mathbb P}}

\def\rR{{\mathbb R}}

\def\indicator{{\mathbbm 1}}

\def\proof{\noindent\textit{Proof:} }
\def\endproof{{\hfill $\clubsuit$ \medskip}}

\title{Bandlimited Signal Reconstruction From the Distribution
of Unknown Sampling Locations\thanks{This work has been supported by grant
no.~P09IRCC039, IRCC, IIT Bombay.}}

\author{\IEEEauthorblockN{Animesh Kumar}
\IEEEauthorblockA{Department of Electrical Engineering\\
Indian Institute of Technology Bombay\\
Mumbai, India -- 400076\\
Email: animesh@ee.iitb.ac.in}}

\begin{document}

\maketitle

\begin{abstract}
We study the reconstruction of bandlimited fields from samples taken at
unknown but statistically distributed sampling locations. The setup is
motivated by distributed sampling where precise knowledge of sensor
locations can be difficult.

Periodic one-dimensional bandlimited fields are considered for
sampling.  Perfect samples of the field at independent and
identically distributed locations are obtained. The statistical
realization of sampling locations is \textit{not known}. First, it is
shown that a bandlimited field cannot be uniquely determined with
samples taken at statistically distributed but unknown locations, even
if the number of samples is infinite.  Next, it is assumed that the
order of sample locations is known. In this case, using insights from
order-statistics, an estimate for the field with useful asymptotic
properties is designed. Distortion (mean-squared error) and 
central-limit are established for this estimate.
\end{abstract}

\IEEEpeerreviewmaketitle

\section{Introduction}
\label{sec:intro}

In the smart-dust paradigm~\cite{kahn99next}, consider a distributed
field sampling problem where sensors are deployed without precise
control on the sensor-locations. One method for distributed field
sampling is to learn the location of these individual sensors, and
then reduce field acquisition to the well-studied non-uniform sampling
problem~\cite{marvastin2001}. However, localization of individual
sensors  in a wireless sensor network can be
difficult~\cite{patwariAKHMCL2005}. In light of these issues, the
reconstruction of a physical field from samples taken at
\textit{unknown} but statistically distributed locations is studied in
this work.

Assuming that the field has a finite support, sensors will have to be
deployed in the finite region where the field is non-zero. The
smoothness of the physical field can be modeled by bandlimitedness. In
this work, it will be assumed that the field is spatially periodic and
bandlimited.  Only \textit{one-dimensional fields} will be considered.
The lack of control in sensor deployment is modeled by a
uniform-distribution on the sensor or sampling-locations.  It is
assumed that sensors are deployed (or scattered) independent of each
other.  Thus, perfect samples of the field at independent and
identically distributed (i.i.d.) but also \textit{unknown} locations
are obtained by the sampling method outlined above.  From these
samples the field has to be estimated. This work focuses on a
\textit{consistent estimate}, that is, an estimate which converges to
the true underlying field when the number of samples is infinite.

The key results shown in this work are as follows:
\begin{enumerate}
\item It will be shown that a bandlimited field \textit{cannot} be
uniquely determined with perfect samples obtained at statistically
distributed locations, even if the number of samples is infinite.
\item If the order of sample locations is known, then using insights
from classical order-statistics, a consistent estimate for the spatial
field is presented.  Distortion (average mean-squared error) and a
central-limit type weak convergence result are established for this
estimate.
\end{enumerate}

\noindent \textit{Prior art:} Recovery of discrete-time bandlimited
signals from samples taken at unknown locations was first studied by
Marziliano and Vetterli~\cite{marzilianoV2000}. Recovery of
a bandlimited signal from a finite number of ordered nonuniform
samples at unknown sampling locations has been studied by
Browning~\cite{browningA2007}.  Estimation of periodic bandlimited
signals in the presence of random sampling location under two models
has been studied by Nordio et al.~\cite{nordioCVP2008}.  Their first
model studies reconstruction of bandlimited signal affected by noise
at random but known locations. Their second model studies estimation
of bandlimited signal from noisy samples on a location set obtained by
random perturbation of equi-spaced deterministic grid.

In contrast, this work presents the estimation of a bandlimited field
from i.i.d.~distributed but unknown samples in an asymptotic setting
(where the number of samples increases to infinity). Asymptotic
consistency (convergence in probability), mean-squared error bounds,
and central-limit type weak law are the focus of this work. The first
key-result of this work is absent in related work due to difference in
the sensing model.


\noindent \textit{Organization:} In Section~\ref{sec:psetup} the field
model, distortion, sensor deployment model, and useful statistical
theory are outlined. In Section~\ref{sec:techresults} asymptotic
consistency, mean-squared error, and weak convergence aspects of field
estimate are discussed. Finally, conclusions will be presented in
Section~\ref{sec:conclusions}.

\section{Problem setup and useful classical results}
\label{sec:psetup}

This section will review the field model, the distortion, and some
useful mathematical results. Field model appears first.

\subsection{Field model and associated properties}
\label{sec:fieldmodel}

The field of interest $g(t)$ is periodic, real-valued, and bandlimited.
Without loss of generality, the period is assumed to be $T = 1$. It is also
assumed that the field $|g(t)| \leq 1$ is bounded.  Bandlimitedness implies
that some $b > 0$ coefficients are non-zero in the Fourier series. Thus,
\begin{align}
g(t) = \sum_{k = -b}^b a_k \exp(j 2 \pi k t ). \label{eq:sigmodel}
\end{align}
Real-valued $g(t)$ implies conjugate symmetry in the Fourier domain, that
is, $a_k = a^*_{-k}$;  however, this symmetry will not be utilized in this
work.  The $(b+1)$ Fourier coefficients constitute the degrees of freedom
for this signal.  With $||g||_\infty \leq 1$, using Bernstein's
inequality~\cite{hardylpi1959}, we get
\begin{align}
|g'(t)| \leq 2 \pi b , \label{eq:bernstein}
\end{align}
where $2\pi b$ rad/s is the bandwidth of the signal. For simplicity of
notation, define $s_b := 1/(2b+1)$ as a spacing parameter and $\phi_k :=
\exp(j 2 \pi k/ (2b+1)) = \exp(j 2 \pi k s_b)$.  By using $(2b+1)$ samples
of the field $g(t)$, its Fourier coefficients can be computed as follows:
\begin{align}
\left[ \begin{array}{c}  g(0) \\ g(s_b) \\ \vdots \\ g(2b s_b) \end{array}
\right]
= \left[ \begin{array}{c c c} 1 &  \ldots & 1 \\ 
\phi_{-b} & \ldots  & \phi_{b} \\
\vdots &  &  \vdots \\ 
(\phi_{-b})^{2b} & \ldots  & (\phi_{b})^{2b} \end{array} \right]
\left[ \begin{array}{c}  a_{-b} \\ a_{-b+1} \\ \vdots \\ a_{b} \end{array}
\right] \nonumber
\end{align}
or more simply
\begin{align}
\vec{g} = \Phi_b \vec{a}, \label{eq:gfroma}
\end{align}
where the vector matrix notation is obvious. The columns of $\Phi_b$ are
orthogonal with a norm-square $(2b+1)$ under the standard inner-product.
The relation in (\ref{eq:gfroma}) is inverted to obtain
\begin{align}
\vec{a} = (\Phi_b)^{-1} \vec{g} = \frac{1}{(2b + 1)} \Phi_b^\dagger
\vec{g}, \label{eq:afromg}
\end{align}
where $\Phi_b^\dagger$ is the conjugate transpose of $\Phi_b$. The
expression in (\ref{eq:afromg}) will be used to obtain an estimate for
$\vec{a}$ as discussed later. 

\subsection{Sensor deployment model and reconstruction distortion}
\label{sec:sensordep}

Denote any sequence as $x_{l}^m := (x_l, x_{l+1}, \ldots, x_m)$ for $m \geq
l$. It will be assumed that sensors are deployed at random locations
$U_1^n$ in the interval of interest $[0,1]$. The locations $U_1^n$ are
i.i.d.~random variables with uniform distribution and probability density
function $f(u) = 1$ for $0 \leq u \leq 1$.  The locations $U_1^n$ are
\textit{not known} in our model. 
An asymptotic number of samples and limiting distribution of $U_1^n$ will
be used for field estimation.
The average mean-squared error will be used as a distortion metric. If
$\estimateG(t)$ is any estimate of $g(t)$, then the distortion is defined
as
\begin{align}
D := \eE(||\estimateG - g||_2^2) := \eE\left[ \int_{0}^{1}  |\estimateG(t)
- g(t)|^2 \mbox{d}t \right]. \label{eq:distortion}
\end{align}
%

\subsection{Useful mathematical results}
\label{sec:conv_results}

For estimation of field from the statistical properties of $U_1^n$,
the following convergence results will be useful. These results for
order-statistics and quantiles are a counterpart to the strong-law of
large numbers (see~\cite[Ch.~10]{davidno2003}).
The ordered version of $U_1^n$ will be denoted by $U_{1:n}^{n:n} :=
\{U_{1:n}, U_{2:n}, \ldots, U_{n:n}\}$ where $U_{n:n}$ is the largest
and $U_{1:n}$ is the smallest~\cite{davidno2003}.

For uniform distribution, the $p$-th population quantile $q_p$ is
equal to $p$. Then with $r = [np] + 1$, it is known
that~\cite[pg.~285]{davidno2003}
\begin{align}
U_{r:n} - p = - (F_n(p) - p) + R_n, \label{eq:quantileU}
\end{align}
where $F_n(u) := \frac{1}{n} \sum_{i = 1}^n \indicator (U_i \leq u) $
is the empirical distribution of $U_1^n$. The remainder term $R_n$ decreases to
$0$ almost surely,
\begin{align}
R_n = O \left( n^{-3/4} (\log n)^{1/2} (\log \log n)^{1/4} \right)
\quad \mbox{as } n \rightarrow \infty. \label{eq:Rn}
\end{align}
By the strong law of large numbers~\cite{durrettp1996}, we know that
$F_n(p) \inas  p$; thus, $U_{r:n} \inas p$ from (\ref{eq:Rn}). 
%
%
Analogous to the central limit theorem, the following fact is
noted.
\begin{fact}\cite[Theorem~10.3]{davidno2003}
\label{fact:weakconv}
Let $0 < p_1 < p_2 < \ldots < p_{2b+1} < 1$ and assume that $(r_i/n -
p_i) = o(1/\sqrt{n}), i = 1,2, \ldots, 2b+1$. Then the following
result holds:
\begin{align}
\sqrt{n}[ U_{r_1:n} - p_1, \ldots, U_{r_{2b+1}:n} - p_{2b+1} ]^T
\weakly {\cal N} \left(\vec{0}, K_U \right), \nonumber
\end{align}
where $[K_U]_{j,j'} = p_j (1 - p_{j'})$ for $j \leq j'$.
%
%
%
%
\end{fact}

All the moments of $U$ are finite since it is bounded (by definition).
The second moment of $U_{r:n} - p$, with $r \approx [np]$ is bounded
by.
\begin{align}
n \eE(U_{r:n} - p)^2 & = p(1-p) \eE(Z^2)  + O(\sqrt{1/n}), \nonumber
\\
& \leq \frac{1}{4} + O(\sqrt{1/n}). \label{eq:mseU}
\end{align}
where $Z \sim {\cal N}(0,1)$ is a normalized Gaussian random variable.

The following fact relates consistency and ${\cal L}^2$ convergence.
\begin{fact}\cite{durrettp1996}
\label{fact:asprop}
If $X_n \inas X$ and $Y_n \inas Y$, then $a X_n + b Y_n \inas a X + b
Y$ for any constants $a, b \in \rR$.  If $X_n$ is bounded and $X_n
\inas X$, then $X_n \inltwo X$.
\end{fact}
We now proceed to the main results of this paper.

\section{Sampling and estimation with random samples}
\label{sec:techresults}

In this section, the key results of this work are presented.  It will be
shown that the field $g(t)$ cannot be inferred uniquely from samples
collected at $U_1^\infty$, where sample-locations are unknown.  Further,
with order information on sample-locations, consistent estimation of the
field is presented.

\subsection{It is impossible to infer $g(t)$ uniquely from $U_1^\infty$}
\label{sec:noorder}

\begin{figure}[!htb]
\begin{center} 
\scalebox{1.0}{\input{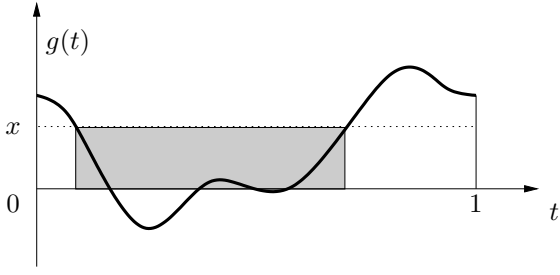}}
\end{center}
\caption{\label{fig:Fgn} \sl \small For any $x \in [-1,1]$ the set of
points where $g(t) \leq x$ can be marked on the $t$-axis. The length or
measure of this set is equal to $\pP(g(U) \leq x)$.}
\end{figure}

It will be shown that if $g(U_1), \ldots, g(U_n)$ is available without the
knowledge of $U_1^n$, then $g(t)$ cannot be inferred uniquely as $n
\rightarrow \infty$. Consider the statistic
\begin{align}
F_{g,n}(x) = \frac{1}{n} \sum_{i = 1}^{n} \indicator(g(U_i) \leq x),
\label{eq:Fgn}
\end{align}
where $\indicator()$ are the indicator random variables. Then $F_{g,n}(x),
x \in [-1, 1]$ completely characterizes the field values $g(U_1), \ldots,
g(U_n)$ and vice-versa. By Glivenko-Cantelli theorem, the right hand limit
in (\ref{eq:Fgn}) converges almost surely to $\pP(g(U) \leq x)$ for all $x \in
[-1,1]$ as $n \uparrow \infty$~\cite{vandervaartA1998}.  This limit is
explained using Fig.~\ref{fig:Fgn}. For any $x \in [-1,1]$ the set of
points where $g(t) \leq x$ can be marked on the $t$-axis. The length or
measure of this set is equal to $\pP(g(U) \leq x)$.

For $0 < \theta < 1$, let $g_\theta(t) = g(t - \theta)$, i.e.,
$g_\theta(t)$ is the shifted version of $g(t)$. Since $g(t)$ is periodic,
its shifts will be cyclic in nature in the period $[0,1]$.  Thus, the
level-sets of $g(t - \theta)$ will be cyclic (in $\theta$) and its measure
$\{u: g_\theta(u) \leq x\}$ will be independent of $\theta$ for every $x
\in [-1,1]$.  Therefore, $\pP(g_\theta(U) \leq x)$ will be independent of
$\theta$ for every $x \in [-1,1]$.
Thus, by only using $F_{g,n}(x)$, which converges to $\pP(g(U) \leq x), x
\in [-1,1]$,  the field $g(t)$ cannot be inferred uniquely. This completes
the discussion of this subsection.

\subsection{Consistent estimation of $g(t)$ from $U_{1:n}^{n:n}$}

From now on, it will be assumed that order information of samples is
available. That is, samples $g(U_{1:n}), \ldots, g(U_{n:n})$ are available
and $g(t)$ has to be estimated.  Using (\ref{eq:afromg}) and the
convergence results in Sec.~\ref{sec:conv_results}, the following estimate
for the Fourier series coefficients of $g(t)$ is proposed:
\begin{align}
\vec{A} := [\estimateA_{-b}, \estimateA_{-b+1}, \ldots, \estimateA_{b}]^T
:= \frac{1}{(2b + 1)} \Phi_b^\dagger \vec{G}.  \label{eq:A_estimate}
\end{align}
where $\vec{G} = [g(U_{1:n}), g(U_{[n \spacing] + 1:n}), \ldots, g(U_{[n 2
b \spacing] + 1:n}) ]^T$. With (\ref{eq:quantileU}) and the smoothness
properties (continuity) of $g(t)$, this estimate is obtained by
substitution method in (\ref{eq:afromg}). Using $\vec{A}$, an estimate for
$g(t)$ is obtained as follows
\begin{align}
\estimateG(t) & = \sum_{k = -b}^b \estimateA_k \exp(j 2 \pi k t ) =
\Phi(t)^T \vec{A} \label{eq:sigest}
\end{align}
where $\Phi(t)^T = \left[ \begin{array}{c c c}
\exp(- j 2 \pi b t) & \ldots & \exp(j 2 \pi bt) 
\end{array} \right]$. Intuitively, $g(t)$ has a finite degrees of freedom.
This enables a procedure to estimate the Fourier series coefficients (the
degrees of freedom) from a finite number of sample estimates of $g(t)$.
Using these estimates of the Fourier series coefficients, the entire field
of interest $g(t)$ can be estimated. For distortion calculation, the
Parseval's theorem~\cite{oppenheimWNS1996} will be useful,
\begin{align}
||\estimateG - g||_2^2 = \sum_{k = - b}^{b} |\estimateA_k - a_k|^2.
\end{align}
A bound on $\eE(|\estimateA_k - a_k|^2)$ will result in a bound on
the expected mean-squared error $\eE(||\estimateG - g||_2^2)$. 

We state our first result now.
\begin{theorem}[Consistency of $\vec{A}$]
\label{thm:conv}
Let $U_{1:n}^{n:n}$ be ordered i.i.d.~Uniform$[0,1]$ random variables.
Define $\vec{A}$ and $\estimateG(t)$ as in (\ref{eq:A_estimate}) and
(\ref{eq:sigest}). Then the estimates $\vec{A}$ and $\estimateG(t)$ are
consistent in almost-sure and ${\cal L}^2$ sense to their respective
limits, i.e.,
\begin{align}
\vec{A} \inas \vec{a}, \widehat{G}(t) \inas g(t) \mbox{ and } \vec{A}
\inltwo \vec{a}, \widehat{G}(t) \inltwo g(t).
\end{align}
\end{theorem}

\proof Only a sketch is provided due to space constraints.  First note that
$U_{[n i s_b] +1:n} \inas i s_b $ for each $i = 0, 1, \ldots, 2b$. Since
$g(t)$ is continuous by assumption, $g(U_{[n i s_b] +1:n}) \inas g(i s_b) $
for each $i = 0, 1, \ldots, 2b$.  Let $\vec{G} := [g(U_{1:n}), g(U_{[n
\spacing] + 1:n}), \ldots, g(U_{[n 2 b \spacing] + 1:n}) ]^T$ and $\vec{g}
:= [g(0), g(s_b), \ldots, g(2b s_b) ]^T$.  By repeated use of
Fact~\ref{fact:asprop}, any finite linear combination $\vec{c}^T \vec{G}$
converges almost-surely to $\vec{c}^T \vec{g}$.  Thus, from
(\ref{eq:A_estimate}), each element of $\vec{A}$ converges almost surely to
$\vec{a}$. Hence, $\vec{A} \inas \vec{a}$.

Next, $\estimateG(t)$ is a finite linear combination of $\vec{A}$.  Since
$\vec{A} \inas \vec{a}$, therefore, $\estimateG(t) \inas g(t)$ in a similar
fashion as above.

For ${\cal L}^2$-convergence, note that $\vec{G}$ is bounded in each
co-ordinate since $|g(t)| \leq 1$ for all $t$. Each element of the matrix
$\Phi_b$ has a magnitude one. Thus, by (\ref{eq:A_estimate}) and the
triangle inequality, $|\estimateA_i| \leq ||g||_\infty \leq 1$ for every $i
= -b, -b + 1, \ldots, b$. Thus, each $\estimateA_i$ is a bounded random
variable. For bounded random sequences, from Fact~\ref{fact:asprop},
$\vec{A} \inas \vec{a}$ implies that $\vec{A} \inltwo \vec{a}$. Similarly,
$|\estimateG(t)| \leq \sum_{k = -b}^{b} |\estimateA_k| \leq (2b+1)$ from
(\ref{eq:sigest}).  Thus, by Fact~\ref{fact:asprop}, $\estimateG(t) \inas
g(t)$ implies $\estimateG(t) \inltwo g(t)$, since $\estimateG(t)$ is
bounded.  \endproof

The second result establishes the scaling of distortion for the proposed
estimate in (\ref{eq:sigest}). 
\begin{theorem}
\label{thm:speed}
Let $U_{1:n}^{n:n}$ be ordered i.i.d.~Uniform$[0,1]$ random variables.
Define $\vec{A}$ and $\estimateG(t)$ as in (\ref{eq:A_estimate}) and
(\ref{eq:sigest}). Then,
\begin{align}
n \eE \left[ ||\estimateG - g||_2^2 \right] \leq \pi^2 b^2 (2b+1) \left[ 1
+ O(\sqrt{1/n}) \right],
\end{align}
that is, the expected distortion decreases as $O(1/n)$.
\end{theorem}

\proof The proof is presented in two parts. First, using the
smoothness properties of $g(t)$, the norm $||\estimateG - g||_2^2$ will
be bounded using the error in quantiles $U_{[np]+1:n} - p$. Next, the
convergence rate of $U_{[np]+1:n} - p$ as in (\ref{eq:mseU}) will be
utilized to upper-bound the distortion.  First note that 
\begin{align}
||\estimateG - g||_2^2 & = \sum_{k = -b}^{b} |\estimateA_k - a_k|^2 \\
& = \frac{1}{(2b+1)^2} ||\Phi_b^\dagger (\vec{G} - \vec{g})||_2^2 \\
& = \frac{1}{(2b+1)^2} \sum_{k = -b}^{b} \left| \sum_{l = 0}^{2b}
[\phi_{k}^{l}]^* (\widehat{G}(l s_b) - g(l s_b))  \right|^2 \nonumber \\
& \stackrel{(a)}{\leq} \frac{(2b+1)}{(2b+1)^2} \sum_{k = -b}^{b} 
\sum_{l = 0}^{2b} |\phi_{k}^{l}| |\widehat{G}(l s_b) - g(l s_b))  |^2
\nonumber \\
& \stackrel{(b)}{=}  \frac{1}{(2b+1)} \sum_{k = -b}^{b} \sum_{l =
0}^{2b} |\widehat{G}(l s_b) - g(l s_b))  |^2 \\
& \stackrel{(c)}{\leq}  \sum_{l = 0}^{2b} |\widehat{G}(l s_b) - g(l
s_b))  |^2.
\end{align}
\begin{align}
& =  \sum_{l = 0}^{2b} |g(U_{[n l s_b] + 1:n}) - g(l s_b))  |^2. \\
& \leq ||g'||_\infty^2 \sum_{l = 0}^{2b} |U_{[n l s_b] + 1:n} - l
s_b|^2. 
\end{align}
where $(a)$ follows by $(a_1 + a_2 + \ldots + a_n)^2 \leq n (a_1^2 +
a_2^2 + \ldots + a_n^2)$, $(b)$ follows by $|\phi_k| = 1$ for all $k$,
and $(c)$ follows since the summation does not depend on $k$. Using
(\ref{eq:mseU}), and taking expectations on both sides
\begin{align}
n \eE \left( || \estimateG - g||_2^2 \right) & \leq  ||g'||_\infty^2
\sum_{l = 0}^{2b} n \eE\left( |U_{[n l s_b] + 1:n} - l s_b|^2\right)
\nonumber \\
& \leq   ||g'||_\infty^2 \sum_{l = 0}^{2b} \left[ \frac{1}{4} +
O(\sqrt{1/n}) \right]
\\
& \leq (2 \pi b)^2 (2b+1) \frac{1}{4} + O(\sqrt{1/n}) \\
& = \pi^2 b^2 (2b+1)[1 + O(\sqrt{1/n})].
\end{align}
This completes the proof.  \endproof

The third result establishes the weak-convergence of $\estimateG(t)$.
\begin{theorem}[Central limit for $\estimateG(t)$]
\label{thm:weaklimit}
Let $U_{1:n}^{n:n}$ be ordered i.i.d.~Uniform$[0,1]$ random variables
and  $\vec{u} = (0, s_b, 2s_b, \ldots 2b s_b)^T$.  Define $\vec{A}$
and $\estimateG(t)$ as in (\ref{eq:A_estimate}) and (\ref{eq:sigest}).
Then the estimate $\vec{A}$ and $\widehat{G}(t)$ satisfy the following
central limits:
\begin{align}
\sqrt{n}(\vec{A} - \vec{a})  \weakly {\cal N} \left(\vec{0},
K_A\right) .
\end{align}
where $K_G = \nabla g^T(\vec{u}) K_U \nabla g (\vec{u})$ and
$K_{\vec{A}}$ is independent of $n$ and given in terms of $K_G$ and $\Phi_b$.  Further,
\begin{align}
\sqrt{n}(\estimateG(t) - g(t)) \weakly {\cal N}\left(\vec{0},
K_G(t)\right). 
\end{align}
where $K_G(t)$ is independent of $n$ and given in terms of
$K_{G}$ and $\Phi_b$.
\end{theorem}

\proof From Fact~\ref{fact:weakconv}, we know that $\vec{U}:=
[U_{1:n}, U_{[n s_b] + 1:n}, \ldots, U_{[n 2b s_b] + 1:n}]^T$ is
asymptotically normal. That is, $\sqrt{n} (\vec{U} - \vec{u}) \weakly
{\cal N}(\vec{0}, K)$, where
\begin{align}
[K]_{i,i'} = (i - 1)s_b [ 1 - (i' - 1) s_b] \mbox{ for } i \leq i'.
\label{eq:Uweak}
\end{align}
Note that $[K]_{i, i'} = [K]_{i', i}$ by the symmetry of a covariance
matrix. Recall $\vec{G}$ from (\ref{eq:A_estimate}). Since $g(t)$ is a
differentiable field, by the delta-method~\cite{vandervaartA1998},
\begin{align}
\sqrt{n} (\vec{G} - \vec{g}) \weakly {\cal N} ( \vec{0}, K_{\vec{G}}),
\end{align}
where $K_{\vec{G}} = \nabla g(\vec{u})^T K \nabla g(\vec{u})$. Observe
that the matrix $K_{\vec{G}}$ depends on the field $g(t)$.  However,
by smoothness of $g(t)$, the vector $\nabla g(\vec{u})$ is bounded and
$K$ is independent of $n$. Thus, $K_{\vec{G}}$ is independent of $n$.
From (\ref{eq:A_estimate}), since $\vec{A}$ is obtained from $\vec{G}$
by a complex-valued linear transformation, we get
\begin{align}
\sqrt{n} (\vec{A} - \vec{a}) \weakly {\cal CN} ( \vec{0},
K_{\vec{A}}).
\end{align}
Observe that the limit is a complex normal Gaussian vector. In
general, the covariance properties of a zero-mean complex random
vector $\vec{S}$ are  determined by $\eE(\vec{S} \vec{S}^\dagger )$
and $\eE(\vec{S} \vec{S}^T)$. Thus, $K_{\vec{A}}$ is determined by the
two matrices $\frac{1}{(2b+1)^2} \Phi_b^\dagger K_{\vec{G}} \Phi_b$
and $\frac{1}{(2b+1)^2} \Phi_b^\dagger K_{\vec{G}} \Phi_b^T$. The
covariance matrix $K_{\vec{G}}$ is independent of $n$; therefore,
$K_{\vec{A}}$ is also independent of $n$ and well defined.

Finally, $\estimateG(t)$ is obtained from $\vec{A}$ by a $t$-dependent
inner product. From (\ref{eq:sigest}), we get $\estimateG(t) =
\Phi(t)^T \vec{A}$. Therefore, $\estimateG(t)$ is a complex normal
Gaussian vector. Its variance can be determined by
$\eE(\estimateG(t)^2)$ and $\eE(|\estimateG(t)|^2)$ which are equal to
$\frac{1}{(2b+1)^2} \Phi(t)^T \Phi_b^\dagger K_{\vec{G}} \Phi_b^T
\Phi(t)$ and $\frac{1}{(2b+1)^2} \Phi(t)^T \Phi_b^\dagger K_{\vec{G}}
\Phi_b \Phi(t)^\dagger$, respectively. Thus the proof is complete. \endproof

This completes our technical result section. The estimation technique
outlined in this section holds well for noise-free setting. If there is
additive noise affecting the samples, then more involved estimation
techniques will be required. Obtaining consistent estimates for $g(l s_b),
l = 0, \ldots, (2b+1)$  is more challenging in the presence of noise.

\section{Conclusions}
\label{sec:conclusions}

The reconstruction of bandlimited fields from samples taken at unknown but
statistically distributed sampling locations was studied. Periodic
one-dimensional bandlimited fields were considered for sampling.  Perfect
samples of the field at i.i.d.~uniform locations were used for the
reconstruction. It was shown that a bandlimited field cannot be uniquely
determined only with samples taken at statistically distributed locations,
even if the number of samples is infinite. Using order information on the
sample locations, a consistent estimate was proposed for the underlying field.
It was shown that this estimate converges in the mean-squared
error sense and almost-sure sense. Further, the mean-squared error
asymptotically decreases as $O(1/n)$, where $n$ is the number of
obtained field samples.


\bibliographystyle{IEEEtran}

\begin{thebibliography}{10}
\providecommand{\url}[1]{#1}
\csname url@samestyle\endcsname
\providecommand{\newblock}{\relax}
\providecommand{\bibinfo}[2]{#2}
\providecommand{\BIBentrySTDinterwordspacing}{\spaceskip=0pt\relax}
\providecommand{\BIBentryALTinterwordstretchfactor}{4}
\providecommand{\BIBentryALTinterwordspacing}{\spaceskip=\fontdimen2\font plus
\BIBentryALTinterwordstretchfactor\fontdimen3\font minus
  \fontdimen4\font\relax}
\providecommand{\BIBforeignlanguage}[2]{{%
\expandafter\ifx\csname l@#1\endcsname\relax
\typeout{** WARNING: IEEEtran.bst: No hyphenation pattern has been}%
\typeout{** loaded for the language `#1'. Using the pattern for}%
\typeout{** the default language instead.}%
\else
\language=\csname l@#1\endcsname
\fi
#2}}
\providecommand{\BIBdecl}{\relax}
\BIBdecl

\bibitem{kahn99next}
\BIBentryALTinterwordspacing
J.~M. Kahn, R.~H. Katz, and K.~S.~J. Pister, ``Next century challenges: Mobile
  networking for ``smart dust'','' in \emph{ACM International Conference on
  Mobile Computing and Networking ({MOBICOM})}, Aug 1999, pp. 271--278.
  [Online]. Available: \url{citeseer.nj.nec.com/kahn99next.html}
\BIBentrySTDinterwordspacing

\bibitem{marvastin2001}
{Farokh Marvasti (ed.)}, \emph{{Nonuniform Sampling}}.\hskip 1em plus 0.5em
  minus 0.4em\relax New York, USA: Kluwer Academic Publishers, 2001.

\bibitem{patwariAKHMCL2005}
N.~Patwari, J.~N. Ash, S.~Kyperountas, A.~O.~H. III, R.~L. Moses, and N.~S.
  Correal, ``Location the nodes: Cooperative localization in wireless sensor
  networks,'' \emph{IEEE Signal Processing Magazine}, vol.~22, no.~4, pp.
  54--69, Jul. 2005.

\bibitem{marzilianoV2000}
P.~Marziliano and M.~Vetterli, ``Reconstruction of irregularly sampled
  discrete-time bandlimited signals with unknown sampling locations,''
  \emph{IEEE Transactions on Signal Processing}, vol.~48, no.~12, pp.
  3462--3471, Dec. 2000.

\bibitem{browningA2007}
J.~Browning, ``Approximating signals from nonuniform continuous time samples at
  unknown locations,'' \emph{IEEE Transactions in Signal Processing}, vol.~55,
  no.~4, pp. 1549--1554, Apr. 2007.

\bibitem{nordioCVP2008}
A.~Nordio, C.-F. Chiasserini, and E.~Viterbo, ``Performance of linear field
  reconstruction techniques with noise and uncertain sensor locations,''
  \emph{IEEE Transactions on Signal Processing}, vol.~56, no.~8, pp.
  3535--3547, Aug. 2008.

\bibitem{hardylpi1959}
G.~H. Hardy, J.~E. Littlewood, and G.~Polya, \emph{Inequalities}.\hskip 1em
  plus 0.5em minus 0.4em\relax London, UK: Cambridge University Press, 1959.

\bibitem{davidno2003}
H.~A. David and H.~N. Nagaraja, \emph{Order Statistics}, 3rd~ed.\hskip 1em plus
  0.5em minus 0.4em\relax New York, NY: John Wiley \& Sons, 2003.

\bibitem{durrettp1996}
R.~Durrett, \emph{Probability: Theory and Examples}, 2nd~ed.\hskip 1em plus
  0.5em minus 0.4em\relax Belmont, CA: Duxbury Press, 1996.

\bibitem{vandervaartA1998}
{A.~W.~van~der~Vaart}, \emph{Asymptotic Statistics}.\hskip 1em plus 0.5em minus
  0.4em\relax Cambridge, UK: Cambridge University Press, 1998.

\bibitem{oppenheimWNS1996}
{Alan Oppenheim and Alan Willsky and Hamid Nawab}, \emph{Signals and Systems},
  2nd~ed.\hskip 1em plus 0.5em minus 0.4em\relax USA: Prentice Hall, 1996.

\end{thebibliography}


\end{document}